\documentstyle[12pt,bezier]{article}
\newfont{\mathea}{msam10 scaled\magstep0}
\newfont{\matheb}{msbm10 scaled 1095}
\newfont{\tmpEins}{cmsy10 scaled 2074}
\newfont{\tmpZwei}{cmsy10 scaled 1095}
\newfont{\tmpDrei}{cmsy10 scaled 1000}
\newfont{\tmpVier}{cmsy5 scaled 1000}
\newfont{\tmpFuenf}{msbm7 scaled\magstep0}

\def\Bbb#1{\mathchoice{\mbox{\matheb #1}}{\mbox{\matheb #1}}%
 {\mbox{\tmpFuenf #1}}{\mbox{\tmpFuenf #1}}}

\def\restriction{\mathchoice{
 \mbox{\unitlength1cm\begin{picture}(.2,.4)%
  \bezier{5}(.07,.3)(.1,.27)(.13,.24)%
  \put(.07,.35){\line(0,-1){.5}}\end{picture}}}{
 \mbox{\unitlength1cm\begin{picture}(.2,.4)%
  \bezier{5}(.07,.3)(.1,.27)(.13,.24)%
  \put(.07,.35){\line(0,-1){.5}}\end{picture}}}{
 \mbox{\mathea\symbol{22}}}{
 \mbox{\mathea\symbol{22}}}}

\def\dach#1#2{\mbox{$\mathop{\vbox{\ialign{%
  $##\crcr\hfil #1 \hfil$\crcr}}}\limits^{\scriptscriptstyle #2}$}}

\def\rnzs{\dach{\rho_2}{\mbox{$\scriptscriptstyle\kern-.7mm0$}}\kern-1.2mm'}

\def\Subset{\mbox{$\subset\kern-.5mm\subset$}}
\newcommand{\LI}{\mbox{{\rm L$^{\kern-.15em\raise.2ex\hbox{\scriptsize 1}}$}}}

\def\Ldummy{\left.\bgroup}
\def\Rdummy{\egroup^{\rule{0mm}{1.4mm}}\right.}
\def\LA{\left\langle\bgroup}
\def\RA{\egroup^{\rule{0mm}{1.4mm}}\right\rangle_{\cal A}^{}}
\def\LR{\left(\bgroup}
\def\RR{\egroup^{\rule{0mm}{1.4mm}}\right)}
\def\LG{\left\{\bgroup}
\def\RG{\egroup^{\rule{0mm}{1.4mm}}\right\}}

\def\Wort#1{\mbox{{\rm #1\kern.1em}}}

\def\lfac#1#2{\vcenter{\hbox{$#1\kern-.2em\raise-.6ex\hbox{\Large{/}}%
 \kern-.2em\raise-1.2ex\hbox{$#2$}$}}}

\def\gin{\mbox{\tmpZwei\symbol{91}\kern-1.4mm\rule{.2mm}{1.85mm}\kern1.4mm}}
\def\gni{\mbox{\tmpZwei\symbol{92}\kern-1.4mm\rule[.15mm]{.2mm}{1.85mm}%
  \kern1.4mm}}

\def\EINS{{\mathop{1\kern-.25em\mbox{{\rm{\small l}}}}}}

\begin{document}

\Large Gamov vectors for Resonances: a Lax-Phillips point of view

\vspace{0.3cm}

\normalsize

H. Baumg\"artel

\vspace{3mm}

Mathematical Institute, University of Potsdam

\vspace{3mm}

Am Neuen Palais 10, PF 601553

\vspace{3mm}

D-14451, Potsdam, Germany

\vspace{3mm}

(e-mail: baumg@rz.uni-potsdam.de)

\vspace{5mm}

Abstract

\vspace{0.3cm}

Results from the Lax-Phillips Scattering Theory are used to analyze quantum mechanical 
scattering systems,
in particular to obtain spectral properties of their resonances which are defined 
to be the poles of the scattering matrix.
For this approach the interplay between the positive energy
projection and the Hardy-space projections is decisive. Among other things 
it turns out that
the spectral
properties of these poles can be described by the (discrete)
eigenvalue spectrum of a so-called truncated evolution, whose eigenvectors 
can be considered as the
Gamov vectors corresponding to these poles. Further an expansion theorem of the
positive Hardy-space part of vectors $Sg$ ($S$ scattering operator) into a series of
Gamov vectors is presented.

\vspace{3mm}


\vspace{1mm}


\section{Introduction}

Hamiltonians $H$ in Quantum Mechanics are semibounded, their absolutely continuous
part is nonnegative in general, the corresponding absolutely continuous spectrum is
the full half line 
$[0,\infty)$
and it is of constant multiplicity.

With regard to scattering problems this leads in many cases to the observation that the 
scattering matrix, if analytically continuable at all, has a cut along the negative
real axis.

On the contrary, the evolutions occurring in the Lax-Phillips (LP-)scattering theory
have generators whose spectrum is pure absolutely continuous, coincides with the real
line and has constant multiplicity, such that also the LP-scattering matrix is defined
on the whole real line as a function of unitary operators on the multiplicity Hilbert
space.

In spite of this contrast the aim of the present paper is to discuss quantum mechanical
scattering from the Lax-Phillips point of view.

We present this approach in two steps. In the first step it is assumed that 
there is no cut. In this case there is a natural
way to connect the quantum mechanical scattering with LP-scattering by an extension
procedure of the Hamiltonians. Then the concepts and methods of the (slightly
generalized) LP-theory can be applied.

In the second step the general case (there is a cut) is explained. The transfer of the 
basic concepts and results from the first step to the general case can be successfully
implemented using the concept {\em pairs of subspaces in generic position}, due to 
Halmos [8]. Decisive results for this topic were given by Kato [9, p.56 ff.] (see
also [2, p.4165 ff.]).

A main result is the spectral characterization of the resonances, i.e. of the
poles of the scattering matrix. This result answers the question where the
eigenvectors of the resonances come from.
It is obtained by the introduction of a {\em truncated evolution} (cf. Skibsted [11],
for example) which is a restriction of a {\em characteristic semigroup} for
$t\geq 0$,
given by the quantum mechanical evolution. The truncated evolution has a pure and discrete
eigenvalue spectrum which is contained in the set of all poles of the scattering matrix
and whose eigenvectors can be interpreted as the {\em Gamov vectors} corresponding to these
poles (for this denotation cf. Bohm/Gadella [5], see also Skibsted [11], Gamov [7]).
Conditions are presented such that every pole of the scattering matrix is an eigenvalue
of the truncated evolution.

The truncated evolution fails to be a semigroup, in general. However, simple conditions 
are presented such that it satisfies the semigroup property.

A second result concerns the expansion of a significant part of vetors 
$Sg$,
($S$ the scattering operator) into a series of Gamov vectors.

\section{Preliminaries}
\subsection{Scattering systems}

A quantum mechanical scattering system
$\{H,H_{0}\}$,
given on a Hilbert space
${\cal H},\,H$
the (selfadjoint) Hamiltonian,
$H_{0}$
the so-called unperturbed Hamiltonian, is called asymptotically complete if the wave
operators
\[
W_{\pm}:=\mbox{s-lim}_{t\rightarrow\pm\infty}e^{itH}e^{-itH_{0}}P^{ac}_{0}
\]
are isometric from
$P^{ac}_{0}{\cal H}$
onto
$P^{ac}{\cal H}$,
where
$P^{ac},P_{0}^{ac}$
are the projections onto the absolutely continuous subspaces of
$H,H_{0}$,
respectively. Then the scattering operator
$S:=W_{+}^{\ast}W_{-}$
is unitary on
$P^{ac}_{0}{\cal H}$
and commutes with the spectral measure of
$H_{0}$.

\subsection{The unperturbed Hamiltonian}

In the following
$P_{0}^{ac}{\cal H}=:{\cal H}_{0}^{+}$
is assumed to be the Hilbert space
${\cal H}_{0}^{+}:=L^{2}([0,\infty),d\lambda,{\cal K})$,
where
${\cal K}$
is a finite-dimensional Hilbert space, describing the multiplicity of the (pure 
absolutely continuous) spectrum of the multiplication operator
$H_{0}^{+}$
on
${\cal H}_{0}^{+}$.
(For example, such Hilbert spaces occur in problems of scattering by a spherically
symmetric potential in
$\Bbb{R}^{3}$,
where
${\cal K}$
denotes the multiplicity space of the angular momentum quantum number
$l\geq 0$.
Also the finite-dimensional Friedrichs model starts with a Hilbert space
${\cal H}:={\cal H}_{0}^{+}\oplus{\cal E}$
where
${\cal E}$
is finite-dimensional.) In this case, due to the commutation property of
$S$
mentioned above,
$S$
is given by an operator function
\[
[0,\infty)\ni\lambda\rightarrow S(\lambda),
\]
a.e. defined, the so-called {\em scattering matrix}, where
$S(\lambda)$
is unitary on
${\cal K}$.

\subsection{The inverse theorem of the scattering theory}

In the following systematic investigation the consequences of special analyticity 
conditions of
$S(\cdot)$
are pointed out. The assumptions, to be presented in the next subsection, are chosen
as a consistent basis for several arguments which are occasionally used in the
resonance framework (see, for example, Bohm/Gadella [5], Gadella [6] and other papers).
To ensure that such additional assumptions on
$S$
do {\em not} imply that there is no Hamiltonian
$H$
such that the scattering system
$\{H,H_{0}\}$
has the scattering operator
$S$
we quote a global existence result, the so-called inverse theorem of the scattering
theory.

\vspace{5mm}

THEOREM 1 (Wollenberg). {\em If}
$S$ 
{\em is unitary on}
${\cal H}_{0}^{+}$
{\em and commutes with the spectral measure of}
$H_{0}^{+}$
{\em then there is always a selfadjoint operator}
$H$
{\em on a Hilbert space}
${\cal H}\supseteq {\cal H}_{0}^{+}$
{\em such that}
$\{H,H_{0}^{+}\}$
{\em is an asymptotically complete scattering system and}
$W_{+}^{\ast}W_{-}=S.$

\vspace{5mm}

We note that
$H$
is not unique of a high degree. However, in the following systematic analysis this fact 
plays no role (see Wollenberg [13] for details, see also Baumg\"artel/Wollenberg
[1, p.240 ff.]).

\subsection{Assumptions on the scattering matrix}

Let
$\Bbb{C}_{<0}:=\{z\in\Bbb{C}: z\neq\lambda,\lambda\leq 0\}$
be the complex plane, cutted by the negative real axis. We assume that
$S(\cdot)$
is analytically continuable into
$\Bbb{C}_{<0}$
with the following properties:

\begin{itemize}
\item[(i)]
$S(\cdot)$
is holomorphic for
$\lambda > 0$,
\item[(ii)]
$S(\cdot)$
is meromorphic on
$\Bbb{C}_{<0}$,
\item[(iii)]
there exist the limits
$\lim_{\epsilon\rightarrow +0}S(\lambda\pm i\epsilon)=:S(\lambda\pm i0)$
for
$\lambda<0$.
\end{itemize}

Note that the unitarity of
$S(\cdot)$
on the positive real axis implies
\[
S(z)^{-1}=S(\overline{z})^{\ast},\quad z\in \Bbb{C}_{<0}.
\]
First implications are

\begin{itemize}
\item[(a)]
$\Bbb{C}_{<0}\ni z\rightarrow S(z)^{-1}$
is also meromorphic , and
$\zeta$
is a pole of
$S(\cdot)$
iff
$\overline{\zeta}$
is a pole of
$S(\cdot)^{-1}$.
\item[(b)]
$S(\lambda-i0)^{\ast}=S(\lambda+i0)^{-1}$
for
$\lambda<0$.
This means that
$S(\lambda+i0)$
is (bounded) invertible for
$\lambda<0$,
but not necessarily unitary.
\item[(c)]
The point
$z=0$
may be a branching point (even of infinite order), but it cannot be a point with pole
character, at most an essential singularity is possible.
\end{itemize}

Concerning the behaviour of
$S(\cdot)$
at infinity we assume

\begin{itemize}
\item[(iv)]
$S(\cdot)$
is bounded at infinity, i.e. there are constants
$C>0,\,R>0$
such that
\[
\Vert S(z)\Vert <C,\quad \vert z\vert>R,\,z\in\Bbb{C}_{<0}.
\]
\end{itemize}

Assumption (iv) ensures maximal transparency and smoothness in the presentation.
However, (iv) is not indispensable (see e.g. 
Bohm/Gadella [5], Gadella [6] and further papers, see also
Strauss [12]).

Simple examples for
$S(\cdot)$
satisfying (i)-(iv) (for the scalar case
${\cal K}:=\Bbb{C}$)
are given by
\[
S(z):=\prod_{j=1}^{r}\frac{z-\overline{\zeta}_{j}}{z-\zeta_{j}},
\]
where the
$\zeta_{1},\zeta_{2},...,\zeta_{r}\in\Bbb{C}$
are nonreal (see e.g. Strauss [12], where Blaschke products are mentioned). Another 
example, where a cut is present, is given by
\[
S(\lambda):=\exp\left\{i\frac{\log\lambda}{\lambda-1}\right\},\quad \lambda>0.
\]
W.r.t. (iv) note that
\[
\vert S(z)\vert\leq\exp\left \{
\frac{\pi\vert x-1\vert+\frac{1}{2}\vert y\vert\log(x^{2}+y^{2})}
{(x-1)^{2}+y^{2}}\right\},\quad z=x+iy.
\]
On the boundary one has
$\vert S(\lambda\pm i0)\vert=\exp\{\mp\frac{\pi}{\lambda-1}\},\,\lambda<0$
and
$\lim\limits_{\lambda\rightarrow-\infty}\vert S(\lambda\pm i0)\vert=1.$
In this case
$z=0$
is an essential (branching) singularity.

\section{The case that there is no cut for the scattering matrix}

If additionally to the assumptions (i)-(iv) the condition

\begin{itemize}
\item[(v)]
$S(\lambda-i0)=S(\lambda+i0),\quad \lambda<0,$
\end{itemize}

\noindent is required then
$S(\lambda+i0)$
is unitary also for
$\lambda<0$,
there is no cut and
$S(\cdot)$
is a unique meromorphic function on
$\Bbb{C}\setminus\{0\}$
 and bounded at infinity. The point
$z=0$
is either a holomorphic one or an essential singularity as the example
$S(z):=\exp(\frac{i}{z})$
shows where
$\vert\exp(\frac{i}{\lambda})\vert=1$
for
$\lambda$
real and
$\lambda\neq 0$.
Moreover
$\exp(\frac{i}{z})\rightarrow 1$
for
$\vert z\vert\rightarrow\infty$.

If
$z=0$
is a holomorphic point then
$S(\cdot)$
is a rational function.
(Note that if (v) is required,
$z=0$
is holomorphic and (iv) is weakened to
"polynomial boundedness at infinity" then still one concludes that
$S(\cdot)$ is rational and
$z=\infty$
is even a holomorphic point.)

We consider the case
"there is no cut"
first because in this case there is a very natural
approach to apply the LP-theory using an extension procedure of the scattering system
in question.
Scattering systems whose scattering matrices are rational are instructive examples 
for this extension procedure.

Later on we introduce a natural transfer of the concepts and results of this case to
the more general case where there is a cut. The crucial method to implement the
transfer is given by the Halmos/Kato results mentioned in the introduction.

\subsection{Extension procedure}

Let
$S(\cdot)$
be a unique analytic function on
$\Bbb{C}\setminus \{0\}$
equipped with the properties (ii) and (iv),
where
$S(\cdot)$
on
$(0,\infty)$
is the scattering matrix 
of an initial scattering system
$\{H^{+},H_{0}^{+}\}$
where
$H^{+}$
is given
on
${\cal H}^{+}\supseteq{\cal H}_{0}^{+}$.
Then 
$\Bbb{R}\setminus\{0\}\ni\lambda\rightarrow S(\lambda)$
is a unitary operator function
on
${\cal K}$.
Then, according to Wollenberg's theorem, there is an appropriate selfadjoint operator
$H$
on a Hilbert space
${\cal H}\supseteq {\cal H}_{0}:=L^{2}(\Bbb{R},d\lambda,{\cal K})$
such that
$\{H,H_{0}\}$
is an asymptotically complete
scattering system,
where now
$H_{0}$
is the multiplication operator on the extended Hilbert space
${\cal H}_{0}$
and
$W_{+}^{\ast}W_{-}=S$,
where
$S$
is the unitary operator on
${\cal H}_{0}$,
given by the unitary operator function
$S(\cdot)$
on
$\Bbb{R}$.
The restriction
$H\restriction E([0,\infty)){\cal H}$,
where
$E(\cdot)$
denotes the spectral measure of
$H$,
together with
$H_{0}\restriction{\cal H}_{0}^{+}=H_{0}^{+}$
yields then an asymptotically complete scattering system with the initial scattering
matrix on
$(0,\infty)$.

Note that
$H\restriction P^{ac}{\cal H}$
is not necessarily an extension of
$H^{+}$
on
$P^{ac}{\cal H}^{+}$
(the absolutely continuous subspace of
$H^{+}$).
However, the scattering operators of the scattering systems
$\{H\restriction E([0,\infty)){\cal H}, H_{0}^{+}\}$
and
$\{H^{+},H_{0}^{+}\}$
coincide (recall that the inverse problem has a vast set of solutions). That is,
the emphasis is only that the extension of the scattering matrix to the whole
real line can be considered as the scattering matrix of a (new) scattering system
w.r.t. the {\em extended} unperturbed Hamiltonian. (Probably under the solutions of 
the inverse problem there is a distinguished one whose restriction coincides with the
initial scattering system, but for the following considerations this has no
relevance.)

\subsection{Example: the one-dimensional perturbation}

Let
${\cal H}_{0},\,H_{0}$
as in Subsection 3.1. Choose
${\cal K}:=\Bbb{C}$
(multiplicity one). Let
$h\in{\cal H}_{0},\,\Vert h\Vert=1$.
Put
\[
H:= H_{0}+P_{h},
\]
where
$P_{h}f:=(h,f)h,\,f\in{\cal H}_{0}$,
is the one-dimensional projection onto
$\Bbb{C}h$.
Obviously,
$\{H,H_{0}\}$
is an asymptotically complete scattering system, the scattering matrix, a scalar,
is given by
\[
S(\lambda)=\frac{\omega(\lambda-i0)}{\omega(\lambda+i0)},\quad \lambda\in\Bbb{R},
\]
where
\[
\omega(z):=1-(h,(z-H_{0})^{-1}h),\quad \mbox{Im}\,z\neq 0.
\]
We choose
\[
h(\lambda):=\pi^{-1/2}\frac{1}{\sqrt{\lambda^{2}+1}}.
\]
Then one obtains
\[
\omega(\lambda\pm i0)=\frac{\lambda-1\pm i}{\lambda\pm i}
\]
and
\[
S(\lambda)=\frac{\lambda^{2}-\lambda +1-i}{\lambda^{2}-\lambda+1+i}.
\]
$H$
is pure absolutely continuous, i.e.
$\{H,H_{0}\}$
on
${\cal H}_{0}$
is an extension of
$\{H\restriction E([0,\infty)){\cal H}_{0},H_{0}^{+}\}$,
where this scattering system plays the role of the initial scattering system.
Note that 
$S(\cdot)$
has a pole 
$\zeta_{0}$
in the upper half plane,
$\zeta_{0}:=\frac{1}{2}\sqrt{5}(-\sin\phi_{0}+i\cos\phi_{0})$
where
$0<\phi_{0}<\frac{\pi}{4}$.

\section{Friedrichs models}

Wollenberg's theorem ensures the (abstract) existence of scattering systems such that 
their scattering matrices satisfy the conditions of Subsection 2.4. In the following 
we present
several Friedrichs models on
$\Bbb{R}$
and
$\Bbb{R}_{+}:=(0,\infty)$
whose scattering matrices realize these conditions. For convenience of the reader
we recall
basic facts on Friedrichs models. We choose the case
$\Bbb{R}_{+}$,
the case
$\Bbb{R}$
is formally similar.

\subsection{The Friedrichs model on $\Bbb{R}_{+}$, general description}

${\cal H}_{0}^{+}$
and
$H_{0}^{+}$
are as before. Let
${\cal E}$
be a finite-dimensional Hilbert space,
$\dim\,{\cal E}<\infty$.
Put
${\cal H}:={\cal H}_{0}^{+}\oplus{\cal E}$.
The projections onto
${\cal H}_{0}^{+}$
and
${\cal E}$
are denoted by
$P_{0},\,P_{\cal E}$,
respectively. Further choose a partial isometry
$\Gamma$ 
with
$\Gamma^{\ast}\Gamma=P_{\cal E}$
and
$\Gamma\Gamma^{\ast}\bot P_{\cal E}$
and a selfadjoint operator
$E_{0}$
on
${\cal E}$
with
$\mbox{spec}\,E_{0}>0.$
Then the (selfadjoint) Hamiltonian is given by
\[
H:= (H_{0}\oplus E_{0})+\Gamma+\Gamma^{\ast}.
\]
The partial isometry
$\Gamma$
can be described by a matrix function
$M(\cdot)$,
a.e. defined, where
$M(\lambda)\in {\cal L}({\cal E}\rightarrow{\cal K})$,
and
${\cal E}\ni e\rightarrow\Gamma e(\lambda)=M(\lambda)e.$
One has
$\int_{0}^{\infty}M(\lambda)^{\ast}M(\lambda)=\EINS_{\cal E}$
and the adjoint 
$\Gamma^{\ast}$
is given by
${\cal H}_{0}\ni f\rightarrow \Gamma^{\ast}f=\int_{0}^{\infty}M(\lambda)^{\ast}
f(\lambda)d\lambda\in{\cal E}.$

Since
$\Gamma+\Gamma^{\ast}$
is finite-dimensional, the wave operators exist and are asymptotically complete, i.e.
\[
W_{\pm}:=\mbox{s-lim}_{t\rightarrow\pm\infty}
e^{itH}e^{-itH_{0}^{+}}P_{0},\quad
W_{\pm}^{\ast}W_{\pm}=P_{0},\quad W_{\pm}W_{\pm}^{\ast}=P^{ac}_{H}.
\] 
We assume that all embedded eigenvalues of
$H_{0}^{+}\oplus E_{0}$
are unstable, i.e.
$P^{ac}_{H}=\EINS_{\cal H}.\; S=W_{+}^{\ast}W_{-}$
is unitary and
$\Bbb{R}_{+}\ni\lambda\rightarrow S(\lambda)$
is a unitary operator function on
${\cal K}$.
The scattering matrix can be calculated using the Liv\v{s}ic-matrix
\[
L_{+}(z):=(z-H_{0}^{+})P_{\cal E}-\Gamma^{\ast}(z-H_{0}^{+})^{-1}\Gamma,\quad
\mbox{Im}\,z>0,
\]
where
\[
\Gamma^{\ast}(z-H_{0}^{+})^{-1}\Gamma=\int_{0}^{\infty}
\frac{M(\lambda)^{\ast}M(\lambda)}{z-\lambda}d\lambda.
\]
Recall that
\[
(L_{+}(z)\restriction{\cal E})^{-1}=P_{\cal E}(z-H)^{-1}P_{\cal E}\restriction{\cal E}.
\]
The right hand side is called the partial resolvent. Therefore, the inverse of the
Liv\v{s}ic-matrix has no poles in
$\Bbb{C}_{+}$.
Moreover, since all embedded eigenvalues are unstable, all poles are contained in
$\Bbb{C}_{-}$.
The scattering matrix is then given by
\[
S(\lambda)=\EINS_{\cal K}-2i\pi M(\lambda)(L_{+}(\lambda+i0)\restriction{\cal E})^{-1}
M(\lambda)^{\ast},\quad \lambda>0.
\]
This shows that the continuability of the scattering matrix into
$\Bbb{C}_{<0}$
depends strongly on the continuability of
$M(\cdot)$.
This will be illustrated by examples. Note that one can distinguish between poles of
$S(\cdot)$
due to the poles of the partial resolvent, i.e. zeros of
$\det(L_{+}(z)\restriction{\cal E})$
in
$\Bbb{C}_{-}$
and poles due to
$M(\cdot)$.
The poles of the partial resolvent are {\em special resonances}. We do not distinguish 
between special resonances and other ones.

\subsection{A Friedrichs model on $\Bbb{R}$ with rational scattering matrix}

The one-dimensional Friedrichs model on
$\Bbb{R}$
is given by the Hamiltonian
$H:=H_{0}+\Gamma+\Gamma^{\ast}$
on the Hilbert space
${\cal H}:={\cal H}_{0}\oplus\Bbb{C}e,\,{\cal K}:=\Bbb{C},\,\Vert e\Vert=1,\,
H_{0}e=\lambda_{0}e,\,\lambda_{0}:=1,$
where
$\Gamma e=:f\in{\cal H}_{0},\,\Vert f\Vert=1,\,\Gamma{\cal H}_{0}=\{0\}.$
Choose
$f(\lambda):=\pi^{-1/2}(\lambda+i)^{-1}.$
Then
\[
(f,(z-H_{0})^{-1}f)=\left\{
\begin{array}{ll}
(z+i)^{-1},& z\in\Bbb{C}_{+}:=\{z\in\Bbb{C}:\mbox{Im}\,z>0\},\\
(z-i)^{-1},& z\in\Bbb{C}_{-}:=\{z\in\Bbb{C}:\mbox{Im}\,z<0\} 
\\
\end{array}
\right.
\]
The Liv\v{s}ic-matrices on
$\Bbb{C}_{+},\Bbb{C}_{-}$
are scalars, given by
\[
\mu_{+}(z)=z-1-\frac{1}{z+i},\quad \mu_{-}(z)=\overline{\mu_{+}(\overline{z})}.
\]
One has
\[
\mu_{-}(\lambda)-\mu_{+}(\lambda)=-\frac{2i}{\lambda^{2}+1}
\]
and the scattering matrix is given by
\[
S(\lambda)=\frac{\mu_{-}(\lambda)}{\mu_{+}(\lambda)}=
\EINS-\frac{2i}{(\lambda-i)(\lambda^{2}-\lambda(1-i)-(1+i))},
\]
$S(\cdot)$ is rational with two poles
$\zeta_{\pm}=\frac{1-i}{2}\pm\sqrt{1+\frac{i}{2}}$
in
$\Bbb{C}_{-}$
and one pole
$\zeta_{0}=i$
in
$\Bbb{C}_{+}$.

That is, if one starts with the scattering system
$\{H\restriction E([0,\infty)){\cal H},H_{0}\restriction{\cal H}_{0}^{+}\}$
with the scattering operator
$S\restriction{\cal H}_{0}^{+}$
then
$H$
realizes an {\em extension} such that
$\{H,H_{0}\}$
is a scattering system with the full scattering operator
$S$.

\subsection{A Friedrichs model on $\Bbb{R}_{+}$ with cut $(-\infty,0)$ for the
scattering matrix}

Choose in Subsection 4.1
${\cal K}:=\Bbb{C},\,{\cal E}:=\Bbb{C}e,\,\Vert e\Vert =1,\,\lambda_{0}=1.$ Then
$M(\cdot)$ 
reduces to a function
$M(\lambda)e=:f(\lambda),\,f\in{\cal H}^{+}_{0}$.
Choose
\[
f(\lambda):=c\frac{\log\lambda}{\lambda-1},\quad \lambda>0,
\]
where 
$c>0$
is the normalizing factor such that
$\Vert f\Vert=1.\,f$
is holomorphic continuable into
$\Bbb{C}_{<0}$
by
$f(z):=c\frac{\log z}{z-1}.$
On the negative real line one has the boundary values
$f(\lambda\pm i0)=c\frac{\log\vert\lambda\vert\pm i\pi}{\lambda-1}$.
Obviously
$f$
is bounded on
$\Bbb{C}_{<0}$.
Putting
\[
g_{+}(z):=\int_{0}^{\infty}\frac{\vert f(\lambda\vert^{2}}{z-\lambda}d\lambda,
\quad z\in\Bbb{C}_{+},
\]
($g_{-}$
on
$\Bbb{C}_{-}$
is defined by the same formula),
the Liv\v{s}ic-matrix on
$\Bbb{C}_{+}$
is given by
\[
\mu_{+}(z)=z-1-g_{+}(z).
\]
On
$\Bbb{C}_{-}$
we have
$\mu_{-}(z)=\overline{\mu_{+}(\overline{z})}.$
On
$\Bbb{R}_{+}$
one has the relation
\[
\mu_{-}(\lambda-i0)-\mu_{+}(\lambda+i0)=-2i\pi\vert f(\lambda)\vert^{2},\quad \lambda>0.
\]
The scattering matrix is given by
\[
S(\lambda)=\frac{\mu_{-}(\lambda-i0)}{\mu_{+}(\lambda+i0)},\quad \lambda>0.
\]
$\mu_{+}$
is holomorphic continuable across
$\Bbb{R}_{+}$
into
$\Bbb{C}_{-}$,
i.e. it is a holomorphic function as is
$\mu_{-}$.
The explicit formulas read
\begin{equation}
\mu_{-}(z)=\mu_{+}(z)-2i\pi\left(\frac{c\log z}{z-1}\right)^{2},\quad z\in\Bbb{C}_{+},
\end{equation}
\begin{equation}
\mu_{+}(z)=\mu_{-}(z)+2i\pi\left(\frac{c\log z}{z-1}\right)^{2},\quad z\in\Bbb{C}_{-}.
\end{equation}
Therefore
$S(\cdot)$
is continuable into
$\Bbb{C}_{<0}$
by
\begin{equation}
S(z):=\frac{\mu_{-}(z)}{\mu_{+}(z)}=1-2i\pi\frac{(c\log z)^{2}}{(z-1)^{2}\mu_{+}(z)},
\quad z\in\Bbb{C}_{<0}.
\end{equation}
For the boundary values on
$\Bbb{R}_{-}:=(-\infty,0)$
we obtain
\[
S(\lambda+i0)=1 - 2i\pi c^{2}\frac{(\log\vert\lambda\vert+i\pi)^{2}}
{(\lambda-1)^{2}\mu_{+}(\lambda+i0)},\quad \lambda<0,
\]
\[
S(\lambda-i0)=\frac{1}{1+2i\pi c^{2}\frac{(\log\vert\lambda\vert-i\pi)^{2}}
{(\lambda-1)^{2}\mu_{-}(\lambda-i0)}},\quad\lambda<0.
\]
Note that
$\mu_{+}(\lambda+i0)=\mu_{-}(\lambda-i0)$
for $\lambda<0$.
A straightforward calculation yields
$S(\lambda+i0)\neq S(\lambda-i0)$
for all
$\lambda<0$,
i.e.
$\Bbb{R}_{-}$
is an actual cut for the scattering matrix.

Furthermore,
$S(\cdot)$
is bounded at infinity. First we show that the functions
$g_{\pm}$
are bounded at infinity. Then formula (2) shows that
$g_{+}$ is bounded at infinity on the whole region
$\Bbb{C}_{<0}$.
Using the decay properties of the integrand at infinity together with Cauchy's theorem
we have
\begin{equation}
c^{2}\int_{-\infty}^{0}\frac{(\log\vert\lambda\vert+i\pi)^{2}}
{(\lambda-1)^{2}(\lambda-z)}d\lambda+g_{+}(z)=2i\pi c^{2}\left(\frac{\log z}{z-1}
\right)^{2},\quad z\in\Bbb{C}_{+},
\end{equation}
(a similar relation is valid for
$g_{-}$
on
$\Bbb{C}_{-}$).
Let
$a>0$
and put
\[
A:=\int_{0}^{a}\frac{\vert f(\lambda\vert^{2}}{\lambda-z}d\lambda,\quad
B:=\int_{a}^{\infty}\frac{\vert f(\lambda\vert^{2}}{\lambda-z}d\lambda.
\]
Then there are constants
$R>0,K>0$
such that
$\vert A\vert<K$
for
$\vert z\vert>R.$
Put
$G:=\{z\in\Bbb{C}:\vert z\vert>R\}.$
Further choose
$0<\delta< a$
and put
$G_{\delta}:=\{z\in\Bbb{C}:\vert z-\lambda\vert\leq\delta,\,a\leq \lambda<\infty\}$.
Then
$\vert B\vert<\frac{1}{\delta}\Vert f\Vert^{2}$
for
$z\in\Bbb{C}\setminus G_{\delta}$.
Then
$A+B$
is bounded on
$G\setminus G_{\delta}$.
Further there is
$0<\gamma< a-\delta$
such that
$\mbox{Re\,z}>\gamma$
for all
$z\in G_{\delta}$
hence a fortiori for all
$z\in G_{\delta}\cap G.$
The formula (4) shows that 
$B$
hence
$A+B$
is also bounded on
$G_{\delta}\cap G$.
That is,
$A+B=g_{+}(\cdot)$
is bounded on
$G$
hence bounded at infinity. The argument for
$g_{-}$
is similar. Since
\[
\vert\mu_{+}(z)\vert\geq\vert z-1\vert-\vert g_{+}(z)\vert >b
\]
with a constant
$b>0$
for sufficiently large
$\vert z\vert$,
we conclude that
$S(\cdot)$
is bounded at infinity.

By a slight modification of this model we can define a Friedrichs model on
$\Bbb{R}$. Put
$f(\cdot)$ 
for
$\lambda>0$
as before and
\[
f(\lambda):=c\frac{\log\vert\lambda\vert+i\pi}{\lambda-1},\quad \lambda<0,
\]
where $c$ is again the (new) normalizing factor.
In this case
$f(\lambda+i0)=f(\lambda),\,f(\lambda-i0)=\overline{f(\lambda)}$
for
$\lambda<0$.
The scattering matrix of this model is (according to Subsection 3.1) given by
\[
S(\lambda)=1-2i\pi\frac{\vert f(\lambda)\vert^{2}}{\mu_{+}(\lambda)},\quad \lambda\in
\Bbb{R}.
\]
In this case the scattering matrix consists of two different analytic functions, defined
on
$\Bbb{R}_{-}$
and
$\Bbb{R}_{+}$.
Starting with the branch on
$\Bbb{R}_{+}$,
by analytic continuation one arrives at the (existing) limits
$S(\lambda\pm i0)$
for
$\lambda<0$,
however these limits are different and have nothing to do with the actual scattering
matrix on the negative half line.

This example shows that Friedrichs models on
$\Bbb{R}$
whose scattering matrix on
$\Bbb{R}_{+}$ have the right properties of analytic continuation are not necessarily
examples for the extension procedure despite of the fact that there is a unitary
scattering matrix on the negative half line.

\section{Extended scattering systems and LP-evolutions}

If there is no cut for the scattering matrix such that the scattering system can
be extended according to 3.1, then a direct connection to the LP-theory can be 
established. However, as the examples in 3.2 and 4.2 show (where there are poles 
in the upper half plane), in general one has to take into account {\em general}
LP-evolutions, where not necessarily outgoing and incoming subspaces are
orthogonal, because this condition excludes poles in the upper half plane (see
Remark 1 in 5.3.2).

\subsection{LP-evolutions}

A unitary strongly continuous evolution group
$\Bbb{R}\ni t\rightarrow U(t):=\exp(-itH)$
on a Hilbert space
${\cal H}$
is called an LP-evolution if there are subspaces
${\cal D}_{+},\,{\cal D}_{-}$
in
${\cal H}$,
called {\em outgoing} and {\em incoming} subspaces, such that
\[
U(t){\cal D}_{+}\subseteq{\cal D}_{+},\,t\geq 0,\quad
U(t){\cal D}_{-}\subseteq{\cal D}_{-},\,t\leq 0,
\]
\[
\bigcap_{t\in\Bbb{R}}U(t){\cal D}_{\pm}=\{0\},\quad
\mbox{clo}\,\{\bigcup_{t\in\Bbb{R}}{\cal D}_{\pm}\}={\cal H},
\]
(see Lax/Phillips [10]). A simple example of an LP-evolution is the so-called (standard)
reference evolution. Let
${\cal H}_{0}:=L^{2}(\Bbb{R},d\lambda,{\cal K})$
as before and
\begin{equation}
(V(t)f)(x):=f(x-t),\quad f\in{\cal H}_{0},
\end{equation}
the regular translation group representation on
${\cal H}_{0}$.
Let
$P_{\pm}$
be the projections given as the multiplication operators by
$\chi_{\Bbb{R}_{\pm}}(\cdot)$,
where
$\chi$
denotes the corresponding characteristic function. Then
$P_{+}{\cal H}_{0},\,P_{-}{\cal H}_{0}$
are outgoing and incoming subspaces for (5), respectively (for details of the
reference evolution see e.g. [1, p.250 ff.].

The {\em spectral representation} of (5) is given by
\[
\hat{V}(t):=FV(t)F^{-1}=e^{-itH_{0}},
\]
where
$F$
denotes the Fourier transformation on
${\cal H}_{0}$:
\[
(Ff)(\lambda):=\frac{1}{\sqrt{2\pi}}\int_{-\infty}^{\infty}e^{-i\lambda x}f(x)dx.
\]
That is, the generator of the spectral representation of the reference evolution is the
"unperturbed Hamiltonian" of an extended scattering system.

Correspondingly, the transformed outgoing/incoming subspaces of the reference evolution
are given by the projections
\[
Q_{\mp}:=FP_{\pm}F^{-1}.
\]
The
$Q_{\pm}$are the projections onto the so-called {\em Hardy spaces}
$Q_{\pm}{\cal H}_{0}=:{\cal H}^{2}_{\pm}$.
For details on Hardy spaces see e.g. [3]. For example, 
$Q_{+}$
is given by
\[
(Q_{+}g)(z)=\frac{1}{2i\pi}\int_{-\infty}^{\infty}\frac{g(\lambda)}{\lambda-z}d\lambda,
\quad g\in{\cal H}_{0},\quad z\in\Bbb{C}_{+}.
\]

\subsection{The evolution
$\Bbb{R}\ni t\rightarrow\exp(-itH)$
for an extended scattering system
$\{H,H_{0}\}$}

\vspace{5mm}

PROPOSITION 1. {\em If}
$\{H,H_{0}\}$
{\em is an extended scattering system then}
$\Bbb{R}\ni t\rightarrow\exp(-itH)$
{\em is an LP-evolution}.

\vspace{3mm}

Proof. One has to define outgoing/incoming subspaces. Let
$W_{\pm}$
denote the wave operators of the scattering system. Define
\[
{\cal D}_{-}:=W_{-}{\cal H}^{2}_{+},\quad {\cal D}_{+}:=W_{+}{\cal H}^{2}_{-}.
\]
The corresponding projections read
\[
D_{-}=W_{-}Q_{+}W_{-}^{\ast},\quad D_{+}=W_{+}Q_{-}W_{+}^{\ast}.
\]
It is an easy calculation to verify the conditions of 3.4.1.
$\Box$

\vspace{3mm}

The representations
\[
W_{+}^{\ast}e^{-itH}W_{+},\quad W_{-}^{\ast}e^{-itH}W_{-},
\]
which both coincide with
$\exp(-itH_{0})$
are called {\em outgoing/incoming spectral representations} of
$\exp(-itH)$, 
respectively. Note that in the outgoing spectral representation
${\cal D}_{+}$
is transformed into
${\cal H}^{2}_{-}$
and in the incoming spectral representation
${\cal D}_{-}$
is transformed into
${\cal H}^{2}_{+}$.

\subsection{The truncated evolution for the extended scattering system}

The spectral properties of the poles of
$S(\cdot)$
become obvious and transparent if one takes into consideration a so-called
{\em truncated evolution}. Its eigenvalue spectrum is contained in the set of all
poles of
$S(\cdot)$
and the corresponding eigenvectors appear as {\em decaying states} w.r.t. the
truncated evolution.

As an essential first step to introduce the truncated evolution we study a characteristic
semigroup for
$t\geq 0$.
It is already introduced in Strauss [12]. This is already a step to give a first
answer to the question where the eigenvectors of the resonances come from.

\subsubsection{The characteristic semigroup}

First we define the characteristic semigroup and present basic properties. We define
an operator
$T(t),\,t\geq 0$
on
${\cal H}$
by
\[
T(t):= D_{+}^{\bot}e^{-itH}D_{+}^{\bot},\quad t\geq 0,
\]
where we put
$D_{+}^{\bot}:=\EINS-D_{+}.$
The transformation of
$T(t)$
into the outgoing spectral representation yields
\[
T_{+}(t):= W_{+}^{\ast}T(t)W_{+}=Q_{+}e^{-itH_{0}}Q_{+},\quad t\geq 0.
\]
It is obvious that
$T_{+}(\cdot)$, hence also
$T(\cdot)$,
is a semigroup for
$t\geq 0$,
because
\[
e^{itH_{0}}Q_{+}=Q_{+}e^{itH_{0}}Q_{+},\quad t\geq 0
\]
(note that
$Q_{+}$
is the incoming projection), hence
\[
Q_{+}e^{-itH_{0}}=Q_{+}e^{-itH_{0}}Q_{+},\quad t\geq 0
\]
and this implies
\[
Q_{+}e^{-it_{1}H_{0}}Q_{+}\cdot Q_{+}e^{-it_{2}H_{0}}Q_{+}=
Q_{+}e^{-i(t_{1}+t_{2})H_{0}}Q_{+},\quad t_{1},t_{2}\geq 0.
\]
Since
$T_{+}(t)$
vanishes on
${\cal H}^{2}_{-}$,
we put
\[
T_{+}(t)\restriction{\cal H}^{2}_{+}=:e^{-itC_{+}},\quad t\geq 0,
\]
where
$C_{+}$
denotes the generator of the restricted semigroup. This semigroup we call the
{\em characteristic semigroup}.

\vspace{5mm}

PROPOSITION 2. {\em The characteristic semigroup}
$T_{+}(\cdot)\restriction{\cal H}^{2}_{+}$
{\em has the following properties}:
\begin{itemize}
\item[(i)]
{\em It is strongly continuous and contractive, the generator}
$C_{+}$
{\em is closed on}
${\cal H}^{2}_{+}$
{\em and}
$\mbox{dom}\,C_{+}$
{\em is dense}.
\item[(ii)]
\[
(T_{+}(t)f)(z)=\frac{1}{2i\pi}\int_{-\infty}^{\infty}
\frac{e^{-it\lambda}}{\lambda-z}f(\lambda)d\lambda,\quad f\in{\cal H}^{2}_{+}.
\]
\item[(iii)]
\[
\mbox{dom}\,C_{+}=\{f\in{\cal H}^{2}_{+}: g_{f}\in{\cal H}^{2}_{+},\quad\mbox{where}\;
g_{f}(z):=zf(z)-\frac{i}{\sqrt{2\pi}}\lim_{x\rightarrow-0}(F^{-1}f)(x)\}
\]
{\em and}
\[
(C_{+}f)(z)=g_{f}(z).
\]
\item[(iv)]
s-$\lim\limits_{t\rightarrow \infty}T_{+}(t)=0.$
\end{itemize}

\vspace{3mm}

A proof can be found in [4]. Second we mention the spectral theory of the characteristic 
semigroup.

\vspace{5mm}

PROPOSITION 3. {\em Let}
$T_{+}(t)\restriction{\cal H}^{2}_{+}=Q_{+}e^{-itH_{0}}\restriction{\cal H}^{2}_{+},\,
t\geq 0,$
{\em as before. Then}
\begin{itemize}
\item[(i)]
$\mbox{res}\,C_{+}=\Bbb{C}_{+}.$
\item[(ii)]
{\em The eigenvalue spectrum of}
$C_{+}$
{\em coincides with}
$\Bbb{C}_{-}$,
{\em i.e. a real point cannot be an eigenvalue}.
\item[(iii)]
{\em The eigenspace of the eigenvalue}
$\zeta\in\Bbb{C}_{-}$
{\em is given by the following subspace}
\[
{\cal N}_{\zeta}:=\{f\in{\cal H}^{2}_{+}: f(z):=\frac{k}{z-\zeta},\,k\in{\cal K}\}.
\]
{\em Then}
\[
T_{+}(t)f=e^{-it\zeta}f,\quad f\in{\cal N}_{\zeta}
\]
{\em follows}.
\end{itemize}

\vspace{3mm}

A proof can be found in [4].

\subsubsection{The truncated evolution for commuting outgoing/incoming projections}

In order to prepare the spectral characterization of the poles of
$S(\cdot)$
we consider the special case that the projections
$D_{+}$
and
$D_{-}$
commute and introduce the truncated evolution in this case.
This special case is an important intermediate step to connect mathematical
(quantum mechanical) scattering theory with LP-ideas for two reasons. First this case
is yet typical LP, because the {\em crucial restriction} (6) of the characteristic
semigroup to the subspace
${\cal H}^{2}_{+}\cap(S{\cal H}_{+}^{2})^{\bot}$
is {\em again} a semigroup, so to say the LP-semigroup in the case in question. Second
already in this case there is a {\em decoupling} of the restriction procedure from
the analytical implications which follow in the case that outgoing and incoming
subspaces are orthogonal (which is a special case of the case of commuting
projections, see Remark 1 in 5.3.2).
The decisive step is a 
modification (resp. further restriction) of the semigroup 
$T(\cdot)$
in the case
$D_{+}D_{-}=D_{-}D_{+}$.
Recall that
\[
T(t)=D_{+}^{\bot}e^{-itH}D_{+}^{\bot}=D_{+}^{\bot}e^{-itH},\quad t\geq 0,
\]
which is due to the relation
$\exp(-itH)D_{+}=D_{+}\exp(-itH)D_{+}$
for
$t\geq 0$,
which is true because
$D_{+}$
is the outgoing projection. We restrict this semigroup further and define
\[
Y(t):=D_{+}^{\bot}e^{-itH}D_{-}^{\bot},\quad t\geq 0.
\]
A straightforward calculation gives
\[
Y(t)=W_{+}Q_{+}e^{-itH_{0}}SQ_{-}W_{-}^{\ast},
\]
i.e. the transformation into the outgoing spectral representation yields
\[
Y_{+}(t)=W_{+}^{\ast}Y(t)W_{+}=Q_{+}e^{-itH_{0}}Q_{+}\cdot SQ_{-}S^{\ast}.
\]

\vspace{5mm}

LEMMA 1. {\em The following relations are equivalent:}
\begin{itemize}
\item[(i)]
$D_{+}D_{-}=D_{-}D_{+},$
\item[(ii)]
$Q_{-}SQ_{+}=SQ_{+}S^{\ast}Q_{-}S,$
\item[(iii)]
$Q_{+}\cdot SQ_{-}S^{\ast}=SQ_{-}S^{\ast}\cdot Q_{+}.$
\end{itemize}
{\em Moreover}
$D_{+}D_{-}=0$
{\em iff}
$Q_{-}SQ_{+}=0.$

\vspace{3mm}

Proof. Straightforward calculation. $\Box$

\vspace{3mm}

That is, the projections
$D_{+}$
and
$D_{-}$
commute iff the projections
$Q_{+}$
and
$SQ_{-}S^{\ast}$
commute. We obtain

\vspace{5mm}

THEOREM 2. {\em If}
$D_{+}$
{\em and}
$D_{-}$
{\em commute then}
$Y_{+}(\cdot)$
{\em hence}
$Y(\cdot)$
{\em is a semigroup for}
$t\geq 0$.

\vspace{3mm}

Proof. We calculate
\begin{eqnarray*}
Y_{+}(t_{1})Y_{+}(t_{2}) &=&
Q_{+}e^{-it_{1}H_{0}}Q_{+}SQ_{-}S^{\ast}Q_{+}e^{-it_{2}H_{0}}Q_{+}SQ_{-}S^{\ast} \\
&=& Q_{+}e^{-it_{1}H_{0}}SQ_{-}S^{\ast}e^{-it_{2}H_{0}}SQ_{-}S^{\ast} \\
&=& Q_{+}Se^{-it_{1}H_{0}}Q_{-}e^{-it_{2}H_{0}}Q_{-}S^{\ast} \\
&=& Q_{+}Se^{-it_{1}H_{0}}e^{-it_{2}H_{0}}Q_{-}S^{\ast} \\
&=& Q_{+}e^{-i(t_{1}+t_{2})H_{0}}Q_{+}\cdot SQ_{-}S^{\ast} \\
&=& Y_{+}(t_{1}+t_{2}). \qquad \Box
\end{eqnarray*}

\vspace{3mm}

Note that 
$Q_{+}\cdot SQ_{-}S^{\ast}$
is the projection of the subspace
$Q_{+}{\cal H}_{0}\cap SQ_{-}{\cal H}_{0}$
hence we obtain
\[
Q_{+}SQ_{-}S^{\ast}{\cal H}_{0}=
Q_{+}{\cal H}_{0}\cap SQ_{-}{\cal H}_{0}=
{\cal H}^{2}_{+}\cap S{\cal H}^{2}_{-}=
{\cal H}^{2}_{+}\cap S({\cal H}^{2}_{+})^{\bot}=
{\cal H}^{2}_{+}\cap (S{\cal H}^{2}_{+})^{\bot}.
\]
This means: the elements of this subspace are exactly those vectors
$f\in{\cal H}^{2}_{+}$
which are orthogonal w.r.t.
$S{\cal H}^{2}_{+}$,
i.e.
$f\bot S{\cal H}^{2}_{+}$.

According to Theorem 2 this subspace is invariant w.r.t. the semigroup
$Y_{+}(\cdot)$. Moreover the semigroup vanishes on the orthogonal complement.
The restriction
\begin{equation}
Z_{+}(t):=Y_{+}(t)\restriction{\cal H}^{2}_{+}\cap(S{\cal H}^{2}_{+})^{\bot},\quad 
t\geq 0
\end{equation}
is a strongly continuous contractive semigroup which is a restriction of the 
characteristic semigroup
$T_{+}(\cdot)\restriction{\cal H}^{2}_{+}.$
This restriction we call the {\em truncated evolution} which is again a semigroup 
in this case.

\vspace{3mm}

REMARK 1. If even
$D_{+}D_{-}=0$,
i.e.
${\cal D}_{+}$
and
${\cal D}_{-}$
are orthogonal then Lemma 1 yields
$Q_{-}SQ_{+}=0$
or, equivalently,
$SQ_{+}=Q_{+}SQ_{+}.$
This means
$S{\cal H}^{2}_{+}\subseteq {\cal H}^{2}_{+}.$
In this case we obtain
\[
{\cal H}^{2}_{+}\cap(S{\cal H}^{2}_{+})^{\bot}={\cal H}^{2}_{+}\ominus
 S{\cal H}^{2}_{+},
\]
i.e. in this case
$Z_{+}(\cdot)$
acts on
${\cal H}^{2}_{+}\ominus S{\cal H}^{2}_{+}$
and it is nothing else than the so-called Lax-Phillips semigroup.Further it turns out that 
in this case
$S(\cdot)$
is holomorphic in
$\Bbb{C}_{+}$
with
$\sup_{z\in\Bbb{C}_{+}}\Vert S(z)\Vert\leq 1$
such that
$S(\lambda)=\mbox{s-lim}_{\epsilon\rightarrow+0}S(\lambda+i\epsilon)$.

\vspace{3mm}

Next we study the spectral theory of the truncated evolution
$Z_{+}(\cdot)$.
Recall that it is a restriction of the characteristic semigroup whose spectral theory
is already known. Therefore, in view of the problem to characterize the eigenvalue 
spectrum of
$Z_{+}(\cdot)$
the crucial question is: Which eigenvalues of the characteristic semigroup, i.e. of
$T_{+}(\cdot)$
on
${\cal H}^{2}_{+}$,
{\em survive} the restriction to the subspace
${\cal H}^{2}_{+}\cap(S{\cal H}^{2}_{+})^{\bot}$?
That is, for
$f_{\zeta,k}\in{\cal N}_{\zeta},\,\zeta\in\Bbb{C}_{-}$,
i.e.
\[
f_{\zeta,k}(\lambda):=\frac{k}{\lambda-\zeta},\quad 0\neq k\in{\cal K},
\]
one has to analyze the condition
\begin{equation}
f_{\zeta,k}\bot S{\cal H}^{2}_{+}.
\end{equation}
Note that (7) means that
$S^{\ast}f_{\zeta,k}\in{\cal H}^{2}_{-}$.
We have
\[
(S^{\ast}f_{\zeta,k})(\lambda)=S(\lambda)^{\ast}f_{\zeta,k}(\lambda)=
\frac{S(\lambda)^{\ast}k}{\lambda-\zeta}=\frac{S(\lambda)^{-1}k}{\lambda-\zeta},
\]
i.e. in any case the vector function
\begin{equation}
\Bbb{C}\ni z\rightarrow(S^{\ast}f_{\zeta,k})(z)=\frac{S(z)^{-1}k}{z-\zeta}
\end{equation}
is meromorphic on
$\Bbb{C}\setminus\{0\}$.
Therefore, a necessary condition for the survival of an eigenvalue
$\zeta\in\Bbb{C}_{-}$
is the following one:
\begin{itemize}
\item[(*)]
There exists
$0\neq k\in{\cal K}$
such that the vector function (8) is holomorphic in
$\Bbb{C}_{-}$.
\end{itemize}

\vspace{5mm}

PROPOSITION 4. {\em If condition} (*) {is satisfied, i.e. there is}
$0\neq k\in{\cal K}$
{\em such that the vector function} (8) {\em is holomorphic in}
$\Bbb{C}_{-}$
{\em then}
$\zeta$
{\em is a pole of}
$S(\cdot)$
{\em or}
$\overline{\zeta}$
{\em is a pole of}
$S(\cdot)$.

\vspace{3mm}

Proof. First, if
$\zeta$
is a holomorphic point for
$S(\cdot)^{-1}$
then one has
$S(\zeta)^{-1}k=S(\overline{\zeta})^{\ast}k=0,$
i.e.
$S(\overline{\zeta})^{\ast}$
is not invertible, which implies that
$\zeta$
is necessarily a pole of
$S(\cdot)$,
because in the contrary one gets
$S(\zeta)=(S(\overline{\zeta})^{\ast})^{-1}$,
a contradiction. 

Second, if
$\zeta$ is a pole of
$S(\cdot)^{-1}$
then
$\overline{\zeta}$
is a pole of
$S(\cdot)$
anyway.
$\quad \Box$

\vspace{5mm}

PROPOSITION 5. {\em Assume that}
$f_{\zeta,k}$
{\em satisfies the condition (*). Then}
$S^{\ast}f_{\zeta,k}\in{\cal H}^{2}_{-}$.

\vspace{3mm}

Proof. Let
${\cal C}_{R}\subset\Bbb{C}$
be the negatively oriented path consisting of the interval
$-R\leq\lambda\leq R$
and the semicircle
${\cal C}_{R,-}:=\{z\in\Bbb{C}_{-}:\vert z\vert=R\}$.
Then for all
$R>0$ we have
\[
\int_{{\cal C}_{R}}\frac{(S^{\ast}f_{\zeta,k})(\lambda)}{\lambda-z}d\lambda=0,
\quad z\in\Bbb{C}_{+}.
\]
Recall that
\[
Q_{+}S^{\ast}f_{\zeta,k}(z)=\frac{1}{2i\pi}\int_{-\infty}^{\infty}
\frac{(S^{\ast}f_{\zeta,k})(\lambda)}{\lambda-z}d\lambda=
\frac{1}{2i\pi}\int_{-\infty}^{\infty}\frac{S(\lambda)^{-1}k}{(\lambda-\zeta)
(\lambda-z)}d\lambda,\,\zeta\in\Bbb{C}_{-},\,z\in\Bbb{C}_{+}.
\]
Further we have
\[
\Vert\int_{{\cal C}_{R,-}}\frac{S(\xi)^{-1}k}{(\xi-\zeta)(\xi-z)}d\xi\Vert \leq
C\Vert k\Vert\int_{{\cal C}_{R,-}}\frac{\vert d\xi\vert}
{\vert \xi-\zeta\vert\cdot\vert\xi-z\vert}\rightarrow 0,\, R\rightarrow\infty.
\]
This implies
\[
\int_{-\infty}^{\infty}\frac{S(\lambda)^{-1}k}{(\lambda-\zeta)(\lambda-z)}d\lambda=0
\]
or
$S^{\ast}f_{\zeta,k}\in{\cal H}^{2}_{-}.\quad \Box$

\vspace{3mm}

According to Propositions 4 and 5 for the survival of the eigenvector
$f_{\zeta,k}$
in the case that
$\zeta\in\Bbb{C}_{-}$, where
$\zeta$
or
$\overline{\zeta}$
is a pole of
$S(\cdot)^{-1}$
one has to analyze the condition (*). At every point
$\zeta\in\Bbb{C}_{-}$
such that
$\zeta$
is a pole of 
$S(\cdot)^{-1}$
or
$\zeta$
is a pole of
$S(\cdot)$
there is a Laurent expansion of
$S(\cdot)^{-1}$
\begin{equation}
S(z)^{-1}=\sum_{n=-m(\zeta)}^{\infty}(z-\zeta)^{n}A_{n,\zeta}.
\end{equation}
which is possibly a power series if
$\zeta$
is a holomorphic point of
$S(\cdot)^{-1}$.

\vspace{5mm}

PROPOSITION 6. {\em Let the coefficients}
$A_{n,\zeta}$
{\em be as in} (9). 
{\em Then }
$S^{\ast}f_{\zeta,k}$
{\em is holomorphic at}
$\zeta$
{\em iff}
\[
A_{n,\zeta}k=0\quad \mbox{for all}\quad n=-m(\zeta),-m(\zeta)+1,...,-1,0.
\]

\vspace{3mm}

Proof. Obvious. $\quad \Box$

\vspace{3mm}

REMARK 2. 
We mention once more the special case 
in the LP-theory, where outgoing and incoming subspaces are orthogonal.
Then
$S(\cdot)$ is holomorphic in
$\Bbb{C}_{+}$
and there are no poles of
$S(\cdot)^{-1}$
in
$\Bbb{C}_{-}$,
i.e. it remains to consider the case where
$\zeta$
is a holomorphic point of
$S(\cdot)^{-1}$,
i.e. it remains the condition
$S(\overline{\zeta})^{\ast}k=0.$
In this case one can also use the following argument: One has
$Su\in{\cal H}^{2}_{+}$
if
$u\in{\cal H}^{2}_{+}$
and one can use the identity
\[
(f_{\zeta,k},Su)=(k,S(\overline{\zeta})u(\overline{\zeta}))_{\cal K},
\]
according to the Paley-Wiener theorem for
$\Bbb{C}_{+}$.

\vspace{3mm}

Under the assumptions of this section the spectrum of the truncated evolution (which is
a semigroup in this case) is a pure eigenvalue spectrum and
${\cal H}^{2}_{+}\cap (S{\cal H}^{2}_{+})^{\bot}$
is spanned by the set of all surviving eigenvectors
$f_{\zeta,k}$.

\subsection{The truncated evolution in the general case}

If
$D_{+}$
and
$D_{-}$
do not commute then the restriction
$Y(\cdot)$
of the semigroup
$T(\cdot)$
resp. its transformation
$Y_{+}(\cdot)$
into the outgong spectral representation fails to be a semigroup. Nevertheless, we
we can restrict 
$Y_{+}(\cdot)$
to
${\cal H}^{2}_{+}\cap(S{\cal H}^{2}_{+})^{\bot}.$
Note that the projection onto this subspace is given by
s-$\lim_{n\rightarrow\infty}(Q_{+}SQ_{-}S^{\ast})^{n}.$
Also in this case we call
\[
T_{+}(t)\restriction{\cal H}^{2}_{+}\cap(S{\cal H}^{2}_{+})^{\bot}=:
Z_{+}(t),\quad t\geq 0,
\]
the {\em truncated evolution} which is not a semigroup. Similarly as before, we can
pose again the question which eigenvalues of the characteristic semigroup
survive this restriction, i.e. we arrive at the same problem, to analyze the condition
$f\bot S{\cal H}^{2}_{+}$,
as before. Therefore, we can transfer the results of Section 3.5.2 (Propositions 4,5
and 6) to the general case. That is, the eigenvalue
$\zeta\in\Bbb{C}_{-}$
and a corresponding eigenvector
$f_{\zeta,k}$
survive the restriction if 
$\zeta$
is a pole of
$S(\cdot)$
or
$S(\cdot)^{-1}$
and
$S^{\ast}f_{\zeta,k}$
is holomorphic in
$\Bbb{C}_{-}$
(see Proposition 6).

In spite of the lack of the semigroup property for the truncated evolution
$Z_{+}(\cdot)$ we obtain for the surviving eigenvalues
$\zeta$
and corresponding eigenvectors
$f_{\zeta,k}$
again 
\[
Z_{+}(t)f_{\zeta,k}=e^{-it\zeta}f_{\zeta,k},\quad t\geq 0,
\]
i.e. restricted to the span of all (surviving) eigenvectors the semigroup property of
$Z_{+}(\cdot)$
remains valid. Note that this span is  finite-dimensional if
$S(\cdot)$
has only finitely many poles.

The eigenvectors of the truncated evolution we call {\em Gamov vectors} (see the
quotations in Section 1).

\section{The case of a cut for the scattering matrix}

If
$(-\infty]$
is an actual cut for the scattering matrix
$S(\cdot)$
on
$\Bbb{C}_{<0}$
then the extension idea does not work because
$S(\cdot)$
cannot be extended to a unitary operator function on the whole real line in a natural way.
Therefore we can work only with the "physical" Hilbert space
${\cal H}_{0}^{+}=P_{+}{\cal H}_{0}$
for the unperturbed Hamiltonian
$H_{0}^{+}=H_{0}\restriction P_{+}{\cal H}_{0}$.

The results of Section 3 suggest that the Hardy spaces, resp. their projections
$Q_{\pm}$
should be crucial concepts also for the case of the existence of a cut. Therefore the
claim is to bring the Hardy spaces into the game in this case. This can be done by 
application of ideas and results of Halmos and Kato to the case of the pairs
$\{P_{+},Q_{+}\}$
and
$\{P_{+},Q_{-}\}$
(see the quotations in Section 1).

\subsection{Pairs of projections in generic position}

DEFINITION. Let
$P,Q$
be projections on a Hilbert space
${\cal H}$.
Then the subspaces
$P{\cal H},Q{\cal H}$
are called {\em subspaces in generic position} if
\[
P{\cal H}\cap Q{\cal H}=
P{\cal H}\cap Q^{\bot}{\cal H}=
P^{\bot}{\cal H}\cap Q{\cal H}=
P^{\bot}{\cal H}\cap Q^{\bot}{\cal H}=\{0\}.
\]

\vspace{2mm}

\noindent Note that for arbitrary projections
$P,Q$
one has
\[
\Vert P-Q\Vert\leq 1
\]
because
$(P-Q)^{2}+(1-P-Q)^{2}=\EINS.$

\vspace{5mm}

THEOREM 3. {\em Let}
$P,Q$
{\em be projections of subspaces in generic position.Then}
\begin{itemize}
\item[(i)]
{\em the linear manifold}
${\cal M}:=PQ{\cal H}\subset P{\cal H}$
{\em is dense in}
${\cal H}$
({\em w.r.t. the Hilbert space topology of}
${\cal H}$).
\item[(ii)]
{\em The projection}
$P$
{\em on}
$Q{\cal H}$
{\em is bijective, i.e. the inverse operator}
$P^{-1}$
{\em exists on}
${\cal M}$.
\item[(iii)]
{\em If}
$\Vert P-Q\Vert<1$
{\em then}
${\cal M}=P{\cal H}$
{\em and}
$P^{-1}$
{\em is continuous on}
$P{\cal H}$.
\item[(iv)]
{\em If}
$\delta:=\Vert(\EINS-P)Q\Vert<1$
{\em then}
$\Vert P-Q\Vert=\delta$.
\item[(v)]
{\em If}
$\Vert P-Q\Vert=1$
{\em then}
$P^{-1}$
{\em is closed and unbounded on}
$P{\cal H}$
{\em and}
$\mbox{dom}\,P^{-1}={\cal M}$
{\em is properly dense in}
$P{\cal H}$.
\end{itemize}

Obviously the pairs
$\{P_{+},Q_{+}\},\,\{P_{+}\,Q_{-}\}$
resp. their corresponding subspaces
\[
\{P_{+}{\cal H}_{0},\,Q_{+}{\cal H}_{0}\},\,\{P_{+}{\cal H}_{0},\,Q_{-}{\cal H}_{0}\}
\]
of
${\cal H}_{0}$
are pairs of subspaces in generic position
as it can be shown easily.
Moreover in this case
${\cal M}_{\pm}:=P_{+}Q_{\pm}{\cal H}_{0}$
is properly dense in
${\cal H}_{0}$.
Therefore all results for such pairs can be used in the 
present context. 
A proof can be found in Kato [9] or in [2]. 
For convenience of the reader we sketch briefly essential arguments.

\vspace{3mm}

Proof of Theorem 3. (i) and (ii): Straightforward calculation. 

(iii): One has
$\delta:=\Vert(\EINS-P)Q\Vert=\Vert Q-PQ\Vert=\Vert(Q-P)Q\Vert\leq\Vert Q-P\Vert<1$.
Let
$f\in Q{\cal H}$. Then
$\Vert f\Vert-\Vert Pf\Vert\leq\Vert f-Pf\Vert=\Vert(\EINS-P)Qf\Vert\leq
\delta\Vert f\Vert$
or
$\Vert Pf\Vert\geq(1-\delta)\Vert f\Vert$,
i.e.
$P^{-1}$
is continuous on
$Q{\cal H}$.

(iv): See Kato [9, p.57]. (v): $\Vert P-Q\Vert=1$ implies $\delta=1.$ Put
${\cal Q}:=Q{\cal H}$
and
$A:=QP^{\bot}Q\restriction{\cal Q}$.
Then
$\mbox{spr}\,A=\Vert A\Vert=\Vert QP^{\bot}P^{\bot}Q\Vert=
\Vert P^{\bot}Q\Vert^{2}=\delta^{2}=1$,
hence
$1\in\mbox{spec}\,A$
follows, but
$1$
is not an eigenvalue of
$A$
because
$Aq=q$
implies
$\mbox{s-lim}_{n\rightarrow\infty}(QP^{\bot})^{n}q=q$,
i.e.
$q\in{\cal Q}\cap P^{\bot}{\cal H}=\{0\}$
and
$\mbox{ker}\,(\EINS_{\cal Q}-A)=\{0\}$.
This means
$(\EINS_{\cal Q}-A)^{-1}$
exists and it is unbounded because
$1\notin\mbox{res}\,A$.
That is,
${\cal D}:=\mbox{dom}\,(\EINS_{\cal Q}-A)^{-1}$
is a proper dense set in
${\cal Q}$
and
$\mbox{ima}\,(\EINS_{\cal Q}-A)={\cal D}=\mbox{ima}\,(Q-QP^{\bot}Q)=
\mbox{ima}\,QPQ$.
The polar decomposition of
$PQ$ reads
$PQ=\mbox{sgn}\,(PQ)\cdot(QPQ)^{1/2}.\quad \mbox{sgn}\,PQ$
maps
$\mbox{ima}\,(QPQ)^{1/2}$
isometrically onto
$\mbox{ima}\,PQ=PQ{\cal H}$
hence
$PQ{\cal H}$
is a proper dense set in
$P{\cal H}$.
That is,
$P\restriction Q{\cal H}$
is unbounded invertible.
$\quad \Box$

\vspace{3mm}

Applying Theorem 3 to the projections
$P_{+},Q_{\pm}$
on
${\cal H}_{0}$
this means:
The projection
$P_{+}$
is a linear bijection of
${\cal H}^{2}_{\pm}$
onto
${\cal M}_{\pm}$
which is properly dense in
$P_{+}{\cal H}_{0},$

\begin{equation}
{\cal H}^{2}_{\pm}\ni f\leftrightarrow f_{+}:=P_{+}f\in{\cal M}_{\pm}=
P_{+}{\cal H}^{2}_{\pm}\subset P_{+}{\cal H}_{0}.
\end{equation}

Therefore we can introduce a new scalar product in
${\cal M}_{\pm}$,
\[
\langle f_{+},g_{+}\rangle :=(f,g),\quad f,g\in{\cal H}^{2}_{\pm}
\]
with the corresponding norm
\begin{equation}
[f_{+}]^{2}:=\Vert f\Vert^{2}=\Vert f_{+}\Vert^{2}+\Vert f_{-}\Vert^{2},
\end{equation}
where
$f_{-}:=P_{-}f,\,P_{-}=\EINS-P_{+},$
i.e. one has
\[
\Vert f_{+}\Vert\leq [f_{+}].
\]
W.r.t. tho (new) norm (11) the linear manifolds
${\cal M}_{\pm}$
are Hilbert spaces
and
because of (11) the linear bijection given by (10) turns out to be isometric.

\subsection{The characteristic semigroup}

Recall that the characteristic semigroup
$T_{+}(\cdot)\restriction {\cal H}^{2}_{0}$
of Section 3 is the transformation of the semigroup
$T(\cdot)\restriction {\cal D}_{+}^{\bot}$
into the outgoing spectral representation (see 3.5.1). By the results of 4.1 now 
it can be transferred to
${\cal M}_{+}$.

\vspace{5mm}

PROPOSITION 7. {\em The assignment}
\begin{equation}
t\rightarrow T_{+}^{P}(t):=P_{+}T_{+}(t)P_{+}^{-1},\quad t\geq 0,
\end{equation}
{\em where}
$P_{+}T_{+}(t)P_{+}^{-1}$
{\em is a linear operator from}
${\cal M}_{+}$
{\em into}
${\cal M}_{+}$,
{\em is a semigroup which is even strongly continuous w.r.t. the (new) Hilbert space
topology of}
${\cal M}_{+}$
{\em and}
$T_{+}^{P}(0)=\EINS_{{\cal M}_{+}}.$

\vspace{3mm}

Proof. Obvious. $\quad\Box$

\vspace{3mm}

The semigroup (12) is the natural counterpart of the characteristic semigroup
in the present case, where there is a cut. The counterpart of
$T(\cdot)\restriction {\cal D}_{+}^{\bot}$
is then given by transformation using the wave operator
$W_{+}$.
Recall that the wave operators
$W_{\pm}$
are defined (and isometric) on
$P_{+}{\cal H}_{0}\supset{\cal M}_{\pm}$,
i.e. the assignment
\begin{equation}
{\cal M}_{+}\ni u\rightarrow W_{+}u\in P^{ac}{\cal H}
\end{equation}
is a bijection, hence we may transfer the (new) Hilbert space norm of
${\cal M}_{+}$
to
$W_{+}{\cal M}_{+}$
by the definition
\begin{equation}
\langle W_{+}f_{+},W_{+}g_{+}\rangle_{W_{+}{\cal M}_{+}}:=
\langle f_{+},g_{+}\rangle,\quad f_{+},\,g_{+}\in{\cal M}_{+}.
\end{equation}
Then the assignment (13) becomes an isometry from
${\cal M}_{+}$
onto
$W_{+}{\cal M}_{+}$
and we obtain

\vspace{5mm}

PROPOSITION 8. {\em The assignment}
\begin{equation}
t\rightarrow W_{+}T_{+}^{P}(t)W_{+}^{\ast},\quad t\geq 0,
\end{equation}
{\em where}
$W_{+}T_{+}^{P}(t)W_{+}^{\ast}$
{\em is a linear operator from}
$W_{+}{\cal M}_{+}$
{\em into}
$W_{+}{\cal M}_{+}$,
{\em is a semigroup which is even strongly continuous w.r.t. the (new)
Hilbert space topology} (14) {of}
$W_{+}{\cal M}_{+}.$

\vspace{3mm}

The proof is obvious. These propositions imply that the spectral theory of the 
characteristic semigroup, developed in 3.5.1, can be completely transferred to
the semigroups (12) and (15).

\vspace{5mm}

COROLLARY 1. {\em The semigroup} (15) {\em resp. its generator has a pure eigenvalue
spectrum which coincides with}
$\Bbb{C}_{-}$.
{\em The eigenspace}
${\cal E}_{\zeta}$
{\em for}
$\zeta\in\Bbb{C}_{-}$
{\em is given by}
${\cal E}_{\zeta}:=W_{+}P_{+}{\cal N}_{\zeta},$
{\em i.e. the eigenvectors are}
$e_{\zeta,k}:=W_{+}P_{+}f_{\zeta,k}$,
{\em where}
\[
P_{+}f_{\zeta,k}(\lambda)=\chi_{[0,\infty)}(\lambda)\frac{k}{\lambda-\zeta}.
\]

\subsection{The truncated evolution}

Recall that in the present case the scattering operator
$S$,
given a priori on
${\cal H}_{0}^{+}$
by the scattering matrix
$\lambda\rightarrow S(\lambda),\,\lambda>0$,
can be extended to a bounded operator on
${\cal H}_{0}$
in two ways by the continuation of
$S(\cdot)$
using the limits of
$S(\cdot)$
on the negative real axis (see 2.4). We put
s-$\lim\limits_{\epsilon\rightarrow\pm 0}S(\lambda\pm i\epsilon)=:S_{\pm}(\lambda)$
for
$\lambda<0$.
Then
$S_{\pm}$
is defined for
$f\in{\cal H}_{0}$
by
\[
(S_{\pm}f)(\lambda)=\left\{
\begin{array}{ll}
S(\lambda)f(\lambda) & \mbox{if}\lambda>0, \\
S_{\pm}(\lambda)f(\lambda) & \mbox{if}\lambda<0.
\\
\end{array}
\right.
\]
Note that
$S_{\pm}$
is not unitary on
${\cal H}_{0}$
but bounded invertible.

Recall further that in 3.5.2 and 3.5.3 the truncated evolution of the characteristic
semigroup is defined by its restriction to the subspace
${\cal H}^{2}_{+}\cap(S{\cal H}^{2}_{+})^{\bot}=
{\cal H}^{2}_{+}\cap S{\cal H}^{2}_{-}.$
In the present case the characteristic semigroup
$T_{+}^{P}(\cdot)$
is defined on
${\cal M}_{+}$.
To define the truncated evolution we have to restrict the characteristic semigroup
to
\[
P_{+}({\cal H}^{2}_{+}\cap S_{\pm}{\cal H}^{2}_{-})\subseteq
P_{+}{\cal H}^{2}_{+}\cap P_{+}(S_{\pm}{\cal H}^{2}_{-})=
P_{+}{\cal H}^{2}_{+}\cap SP_{+}{\cal H}^{2}_{-}={\cal M}_{+}\cap S{\cal M}_{-},
\]
i.e. the restriction is independent of the ambiguity that the continuation of the
scattering matrix to the negative half line is not unique.
The truncated evolution is then defined by

\begin{equation}
t\rightarrow T_{+}^{P}(t)\restriction P_{+}({\cal H}^{2}_{+}\cap
S_{\pm}{\cal H}^{2}_{-}),
\quad t\geq 0.
\end{equation}

If
${\cal H}^{2}_{+}\cap S_{\pm}{\cal H}_{-}\supset\{0\}$
then also
$P_{+}({\cal H}^{2}_{+}\cap S_{\pm}{\cal H}^{2}_{-})\supset\{0\}.$

For example, if in the "no cut case"
$D_{+}D_{-}=0$,
i.e.
$S{\cal H}^{2}_{+}\subseteq{\cal H}^{2}_{+}$,
and
${\cal H}^{2}_{+}\cap S{\cal H}^{2}_{-}=\{0\},$
such that
$S{\cal H}^{2}_{+}={\cal H}^{2}_{+}$
follows then one obtains that
$S(\cdot)$
has no poles in
$\Bbb{C}\setminus\{0\}$.
In the "rational case" this means
$S=\EINS$.

The spectral theory of (16) can be immediately traced back to that of the truncated
evolution in  3.5.3 (and 3.5.2, in particular see Proposition 6).

\vspace{5mm}

COROLLARY 2. {\em The eigenvalue}
$\zeta\in\Bbb{C}_{-}$
{\em and a corresponding eigenvector}
$P_{+}f_{\zeta,k}$
{\em of the characteristic semigroup}
$T_{+}^{P}(\cdot)\restriction {\cal M}_{+}$
{\em survive the restriction} (16) {\em iff}
$\zeta$
{\em is a pole of}
$S(\cdot)$
{\em or}
$S(\cdot)^{-1}$
{\em and}
$S^{\ast}f_{\zeta,k}$
{\em is holomorphic in}
$\Bbb{C}_{-}.$

\vspace{3mm}

Proof. The condition for survival of
$P_{+}f_{\zeta,k}$
reads
$S^{-1}P_{+}f_{\zeta,k}\in{\cal M}_{-}$
or
$P_{+}^{-1}S^{-1}P_{+}f_{\zeta,k}\in{\cal H}^{2}_{-}.$
Now
\[
(S^{-1}P_{+}f_{\zeta,k})(\lambda)=\chi_{[o,\infty)}(\lambda)\frac{S(\lambda)^{-1}k}
{\lambda-\zeta}
\]
and
\[
(P_{+}^{-1}S^{-1}P_{+}f_{\zeta,k})(\lambda)=
\frac{S(\lambda)^{-1}k}{\lambda-\zeta}.
\]
That is, we arrive at the same conditions as in 3.5.2.
$\quad \Box$

\section{Expansion of vectors from
$Q_{+}S_{-}{\cal H}^{2}_{-}$
in a series of Gamov vectors}

In this section we require that
$S(\cdot)$
has only finitely many poles in
$\Bbb{C}_{<0}$
(cf. condition (ii) in 2.4).
First recall that
$P_{+}{\cal H}^{2}_{-}$
is dense in
$P_{+}{\cal H}_{0}$,
i.e. each vector
$u\in P_{+}{\cal H}_{0}$
can be approximated by a vector
$P_{+}g$
where 
$g\in{\cal H}^{2}_{-}$
such that
$\Vert u-P_{+}g\Vert$
is arbitrary small. Then an expansion of
$Q_{+}SP_{+}g$
is an approximation for the vector
$Q_{+}Su$.

\vspace{5mm}

THEOREM 4. {\em Let}
$g\in{\cal H}^{2}_{-}$
{\em and}
$S_{-}$
{\em be the operator of} 4.3. {\em Then there is an expansion of}
$Q_{+}S_{-}g$
{into a series of Gamov vectors}
\[
(Q_{+}S_{-}g)(z)=\sum_{j=1}^{r}\frac{S_{-1,j}g(\zeta_{j})}{z-\zeta_{j}},
\quad z\in \Bbb{C}_{+},
\]
{\em where the}
$\zeta_{j}\in\Bbb{C}_{-}$
{\em run through all poles of}
$S(\cdot)$
{\em in}
$\Bbb{C}_{-}$
{\em and where}
$S_{-1,j}$
{\em denotes the residuum of}
$S(\cdot)$
{\em at}
$\zeta_{j}.$

\vspace{3mm}

Proof. Let
$f\in{\cal H}^{2}_{+}$
be arbitrary and put
\[
F(z):=(f(\overline{z}),S(z)g(z))_{\cal K},\quad z\in \Bbb{C}_{-}\cup\Bbb{R}.
\]
Consider the positively oriented path
${\cal C}$
in
$\Bbb{C}$
consisting of two pieces
${\cal C}={\cal C}_{0}\cup{\cal C}_{1}$,
where
${\cal C}_{0}:=[-R,R],\,R>0$
and
${\cal C}_{1}$
consists of the following three segments:
\[
{\cal C}_{1,1}:=\{-R-iy:\,0\leq y\leq\delta\},
\]
\[
{\cal C}_{1,2}:=\{x-i\delta:\,-R\leq x\leq R\},
\]
\[
{\cal C}_{1,3}:=\{R-iy:\,0\leq y\leq\delta\},
\]
where
$\delta>0$.
Then
\[
\int_{\cal C}F(z)dz=-\int_{-R}^{R}F(\lambda)d\lambda +
\int_{{\cal C}_{1}}F(z)dz=2i\pi\mbox{Res}_{\Bbb{C}_{-}}F(\cdot),
\]
if
$R$
and
$\delta$
are sufficiently large. The integral
$\int_{{\cal C}_{1}}$
is estimated as follows, according to the adaption of an argument of 
Yosida [14, p.163 ff.]. First we have
\[
\left\vert\int_{{\cal C}_{1}}(f(\overline{z}),S(z)g(z))_{\cal K}dz\right\vert\leq
C\int_{{\cal C}_{1}}\Vert f(\overline{z})\Vert\cdot\Vert g(z)\Vert\cdot
\vert dz\vert\leq
C\left(\int_{{\cal C}_{1}}\Vert f(\overline{z})\Vert^{2}\vert dz\vert\right)^{1/2}
\left(\int_{{\cal C}_{1}}\Vert g(z)\Vert^{2}\vert dz\vert\right)^{1/2}.
\]
We consider the first integral (the second one can be treated similarly). The
complex conjugated path
$\overline{{\cal C}_{1}}$
consists of three parts, so we have to estimate
\begin{equation}
\int_{0}^{\delta}\Vert f(-R+iy)\Vert^{2}dy+\int_{-R}^{R}
\Vert f(x+i\delta)\Vert^{2}dx+\int_{0}^{\delta}
\Vert f(R+iy)\Vert^{2}dy=A+B+C.
\end{equation}
First note that
\begin{equation}
\int_{-\infty}^{\infty}\Vert f(x+i\delta)\Vert^{2}dx=\int_{0}^{\infty}
e^{-2p\delta}\Vert \hat{f}(-p)\Vert^{2}dp,
\end{equation}
where
$\hat{f}$
is the Fourier transform of
$f$. (18) shows that the integral on the left hand side is sufficiently small
if
$\delta$
is large enough.

Second, to each
$\delta>0$
there is a sequence
$R_{n}\rightarrow\infty$
such that
\[
\lim_{n\rightarrow\infty}\int_{0}^{\delta}\Vert f(\pm R_{n}+iy)\Vert^{2}dy=0,
\]
because in the contrary case there is
$\delta_{0}>0$
and
$R_{0}>0$
and a constant
$b>0$
such that for all
$\vert x\vert\geq R_{0}$
\[
\int_{0}^{\delta}\Vert f(x+iy)\Vert^{2}dy\geq b
\]
hence
\[
\int_{\vert x\vert\geq R_{0}}\{\int_{0}^{\delta}\Vert f(x+iy)\Vert^{2}dy\}dx=\infty
\]
which is a contradiction to
\[
\int_{-\infty}^{\infty}\{\int_{0}^{\delta}\Vert f(x+iy)\Vert^{2}dy\}dx=
\int_{0}^{\delta}\{\int_{-\infty}^{\infty}\Vert f(x+iy)\Vert^{2}dx\}dy<\infty.
\]
Now choose first
$\delta>0$ 
large such that $B$ is small (uniformly for all
$R>0$).
To this
$\delta$
there corresponds a sequence
$R_{n}\rightarrow\infty$.
Then
$\lim\limits_{n\rightarrow\infty}A=\lim\limits_{n\rightarrow\infty}C=0$,
this implies that (17) is sufficiently small and we obtain
\[
\lim_{R\rightarrow\infty,\delta\rightarrow\infty}\int_{{\cal C}_{1}}F(z)dz=0
\]
and
\[
\int_{-\infty}^{\infty}(f(\lambda+i0),S_{-}(\lambda)g(\lambda-i0))_{\cal K}d\lambda=
-2i\pi\mbox{Res}_{\Bbb{C}_{-}}F(\cdot).
\]
Now the coefficient of
$(z-\zeta)^{-1}$
in the Laurent expansion of
$F(\cdot)$
at
$\zeta_{j}$
reads
$(f(\overline{\zeta_{j}}),S_{-1,j}g(\zeta_{j}))_{\cal K}$,
where
$S_{-1,j}$
is that coefficient for
$S(\cdot)$.
Then we obtain
\[
\mbox{Res}_{\Bbb{C}_{-}}\,F(\cdot)=\sum_{j=1}^{r}
(f(\overline{\zeta_{j}}),S_{-1,j}g(\zeta_{j}))_{\cal K}.
\]
Now a well-known Hardy space theorem says that
\[
-(f(\overline{\zeta_{j}}),S_{-1,j}g(\zeta_{j}))_{\cal K}=
\frac{1}{2i\pi}\int_{-\infty}^{\infty}
\left(f(\lambda+i0),\frac{S_{-1,j}g(\zeta_{j})}{\lambda-\zeta_{j}}\right)_{\cal K}d\lambda.
\]
This gives
\[
\int_{-\infty}^{\infty}(f(\lambda+i0),S_{-}(\lambda)g(\lambda-i0))_{\cal K}d\lambda=
\int_{-\infty}^{\infty}\left(f(\lambda+i0),
\sum_{j=1}^{r}\frac{S_{-1,j}g(\zeta_{j})}{\lambda-\zeta_{j}}\right)_{\cal K}d\lambda
\]
or
\[
\int_{-\infty}^{\infty}\left(f(\lambda+i0),S_{-}(\lambda)g(\lambda-i0)-
\sum_{j=1}^{r}\frac{S_{-1,j}g(\zeta_{j})}{\lambda-\zeta_{j}}\right)_{\cal K}d\lambda=0.
\]
Since
$f\in{\cal H}^{2}_{+}$
is arbitrary this implies that the function
\[
\Bbb{R}\ni\lambda\rightarrow S_{-}(\lambda)g(\lambda)-
\sum_{j=1}^{r}\frac{S_{-1,j}g(\zeta_{j})}{\lambda-\zeta_{j}}
\]
is orthogonal to
${\cal H}^{2}_{+}$
hence an element of
${\cal H}^{2}_{-}$.
However, the part
\[
\lambda\rightarrow\sum_{j=1}^{r}\frac{S_{-1,j}g(\zeta_{j})}{\lambda-\zeta_{j}}
\]
is from
${\cal H}^{2}_{+}$.
This yields the assertion.
$\quad \Box$

\vspace{5mm}

REMARK 3. From the proof of Theorem 4 we extract the relation

\begin{eqnarray*}
(f,S_{-}g) &=&
\int_{-\infty}^{0}(f(\lambda+i0),S_{-}(\lambda)g(\lambda-i0))_{\cal K}d\lambda+
\int_{0}^{\infty}(f(\lambda+i0),S(\lambda)g(\lambda-i0))_{\cal K}d\lambda \\
&=& -2i\pi\sum_{j=1}^{r}(f(\overline{\zeta_{j}}),S_{-1,j}g(\zeta_{j}))_{\cal K}.
\end{eqnarray*}

Since the physical transition probability from the state
$P_{+}g$
to the state
$P_{+}f$
is given by the modulus square of
\[
(P_{+}f,P_{+}g)=\int_{0}^{\infty}(f(\lambda+i0),S(\lambda)g(\lambda-i0))_{\cal K}d\lambda,
\]
we obtain
\[
(P_{+}f,P_{+}g)=-2i\pi\sum_{j=1}^{r}(f(\overline{\zeta_{j}}),
S_{-1,j}g(\zeta_{j}))_{\cal K}-\int_{-\infty}^{0}
(f(\lambda+i0),S_{-}(\lambda)g(\lambda-i0))_{\cal K}d\lambda,
\]
i.e. this term is given by the sum of a {\em residual term} and a so-called
{\em background integral} due to the "virtual" negative energies.

\section{Conclusions}
The presented results suggest their application to scattering systems with embedded
eigenvalues of the unperturbed Hamiltonian, e.g. Friedrichs models with
{\em special} resonances (nonreal zeros of the determinant
of their Liv\v{s}ic-matrix (see Section 4, see also [15]). Note that for these resonances 
eigenfunctionals
w.r.t. the Hamiltonian can be constructed (see [15] and [16]). It would be nice 
to clarify the connection between these eigenfunctionals and the Gamov vectors in 
that case.

\section{Acknowledgments}

It is a pleasure to thank Professors A. Bohm and M. Gadella for discussions on the 
subject at the 
conference on "Irreversible Quantum Dynamics" in Trieste, 29th July - 2th August 2002,
Professor A. Bohm for dicussions at the CFIF-Workshop on "Time Asymmetric
Quantum Theory: The Theory of Resonances", 23th - 26th July 2003, Lisbon and Professor
Y. Strauss for discussions at the 25th International Colloquium on
Group Theoretical Methods in Physics
in Cocoyoc, Mexico,
2th - 6th August 2004.

\section{References}

\noindent [1] Baumg\"artel, H., Wollenberg, M.: Mathematical Scattering Theory, 
Birkh\"auser Basel Boston Stuttgart 1983

\vspace{3mm}

\noindent [2] Baumg\"artel, H., Jurke, M., Lled\`{o}, F.: Twisted duality of the
CAR-algebra, J. Math. Phys. 43(8), 4158-4179 (2002)

\vspace{3mm}

\noindent [3] Baumg\"artel, H.: Introduction to Hardy spaces, Internat. J. of
Theor. Phys. 42, No.10, 2211-2221 (2003)

\vspace{3mm}

\noindent [4] Baumg\"artel, H.: On Lax-Phillips semigroups,
to be published

\vspace{3mm}

\noindent [5] Bohm, A, Gadella, M.: Dirac Kets, Gamov vectors and Gelfand Triplets,
Lecture Notes in Physics 348, Springer Verlag 1989

\vspace{3mm}

\noindent [6] Gadella, M.: A rigged Hilbert space of Hardy class functions:
Application to resonances, J. Math. Phys. 24(6), 1462-1469 (1983)

\vspace{3mm}

\noindent [7] Gamov, G.: Zur Quantentheorie des Atomkernes, Z. Phys. 51, 204-212 (1928)

\vspace{3mm}

\noindent [8] Halmos, P.R.: Two subspaces, Trans. Amer. Math. Soc. 144,
381-389 (1969)

\vspace{3mm}

\noindent [9] Kato, T.: Perturbation Theory for Linear Operators, Springer Verlag 
Berlin 1976

\vspace{3mm}

\noindent [10] Lax, P.D., Phillips, R.S.: Scattering Theory, Academic Press, New York
1967

\vspace{3mm}

\noindent [11] Skibsted, E.: Truncated Gamov Functions, $\alpha$-decay and
the Exponential Law, Commun. Math. Phys. 104, 591-604 (1986)

\vspace{3mm}

\noindent [12] Strauss, Y.: Resonances in the Rigged Hilbert Space and Lax-Phillips
Scattering Theory, Internat. J. of Theor. Phys. 42, No.10, 2285-2317 (2003)

\vspace{3mm}

\noindent [13] Wollenberg, M.: On the inverse problem in the abstract theory of
scattering, ZIMM-Preprint Akad. Wiss. DDR, Berlin 1977

\vspace{3mm}

\noindent [14] Yosida, K.: Functional Analysis, Springer Verlag Berlin 1971

\vspace{3mm}

\noindent [15] Baumg\"artel, H.: Resonances and Virtual Poles in Scattering
Theory, Internat, J. of Theor. Phys. 42, No.10, 2379-2388 (2003)

\vspace{3mm}

\noindent [16] Baumg\"artel, H.: Resonances of Perturbed Selfadjoint Operators and 
their Eigenfunctionals, Math. Nachr. 75, 133-151 (1976)

\section{Addendum}

Extended and refined version of a talk presented at the 25th International
Colloquium on Group Theoretical Methods in Physics, Section Semigroups,
Time Asymmetry, and Resonances, in Cocoyoc, Mexico, 2th - 6th August 2004.

\end{document}